\documentclass[showpacs,superscriptaddress,floatfix,reprint,prb]{revtex4-1}
\usepackage[latin1]{inputenc}
\usepackage[english]{babel}
\usepackage{graphicx}
\usepackage{amsmath}
\usepackage{amssymb}
\usepackage{amscd}
\usepackage{eucal}
\usepackage{color}
\usepackage{bm}
\usepackage{xfrac}
\def\vdelta{u}
\input {epsf}

\begin{document}

\title{Effects of electron scattering on the topological properties of
nanowires: Majorana fermions from disorder and superlattices}

\author{\.{I}. Adagideli}
\affiliation{Faculty of Engineering and Natural Sciences, Sabanci University,
Orhanli-Tuzla, Istanbul, Turkey}
\author{M. Wimmer}
\affiliation{Instituut-Lorentz, Universiteit Leiden, P.O. Box 9506, 2300 RA Leiden, The Netherlands}
\author{A. Teker}
\affiliation{Faculty of Engineering and Natural Sciences, Sabanci University,
Orhanli-Tuzla, Istanbul, Turkey}
\date{\today }

\begin{abstract}
We focus on inducing topological state from regular, or irregular
scattering in (i) p-wave superconducting wires and (ii) 
Rashba wires proximity coupled to an s-wave superconductor. 
We find that contrary to common expectations the topological properties of both systems are fundamentally different:
In p-wave wires, disorder generally has a detrimental effect on the topological order and the topological state is destroyed
beyond a critical disorder strength. In contrast, in Rashba wires, which are 
relevant for recent experiments, disorder can {\it induce} topological order, reducing the need for 
quasiballistic samples 
to obtain Majorana fermions.
Moreover, we find that the total phase space area of the topological state is conserved for long disordered Rashba wires,
and can even be increased in an appropriately engineered superlattice potential. 
\end{abstract}


% 74.78.Na 	Mesoscopic and nanoscale systems
% 74.20.Mn 	Nonconventional mechanisms
% 74.45.+c 	Proximity effects; Andreev reflection; SN and SNS junctions
% 71.23.-k 	Electronic structure of disordered solids

\maketitle

\section{Introduction} 

The response of conventional s-wave superconductors to nonmagnetic disorder is drastically different from that of 
nonconventional superconductors with higher angular momentum pairing. While s-wave superconductivity is resistant 
to the presence of nonmagnetic disorder, it is detrimental to unconventional 
superconductivity~\cite{Anderson59, Abrikosov60}.  Other than the pairing symmetry, superconductors are also classified
by the structure of their quasiparticle excitations: those that can be adiabatically transformed into a conventional 
insulator are topologically trivial. The topologically nontrivial superconductors on the other hand
are distinguished by exotic low-energy excitations at their
boundaries. In one dimension, these excitations turn out to be their
own antiparticles and are dubbed Majorana 
fermions. Thus, Majorana fermions can appear 
at the ends of a spinless p-wave superconducting wire~\cite{Kitaev01}
or at the ends of a spin-orbit coupled semiconductor quantum wire in proximity to a conventional s-wave 
superconductor~\cite{Lutchyn10,Oreg10}. 

 The latter, hybrid system reduces to an effective p-wave
superconductor~\cite{Alicea11} in the limit of an almost depleted
wire. For this reason, the topological properties of p-wave
superconducting wires and hybrid nanowire systems with s-wave
superconductivity are commonly assumed to be equivalent~\cite{Alicea12,
  Leijnse12}.  In particular, the effects of disorder on the
topological superconductivity (and thus on the Majorana fermion) have
so far been explored mainly within this premise.
The main conclusion 
of these works is that disorder is {\it always} detrimental to the topological superconductivity and 
hence the Majorana fermion can survive only if (i)~the mobility is high enough such that the localization
length is longer than the coherence length of the topological
superconductor~\cite{Piet11, Piet11a, Potter11, Sau12, Maresa12} and (ii)~there is an odd number of spin-resolved 
transverse modes in a multi-mode wire~\cite{Wimmer10,Potter10,Stanescu11,Piet12}. 

The recent observation of a zero-bias 
peak (ZBP) in the Andreev conductance
of superconducting InSb nanowire heterostructures~\cite{mourik12}, followed by similar observations
subsequentially reported by other groups~\cite{Deng12, Das12},
therefore raised many questions
about the origin of the peak because the mean free path obtained from normal state 
conductance shows the nanowires to be too dirty to be in the topological regime.
Indeed, recent works caution 
against the interpretation
that these peaks are signatures of Majorana fermions \cite{Liu12,Bagrets12,Pikulin12, Rainis13}.

In contrast, here we show that topological superconductivity in the presence of s-wave order parameter is resistant to
disorder in that the conditions (i) and (ii) are in
fact not essential for the survival of Majorana fermions. 
The underlying reason  (which is not captured by
an effective p-wave model)
is that a transport gap can be utilized to induce and protect
topological state similar to the spectral gaps of conventional proposals. 
Hence, disorder can induce robust topological order in s-wave
superconductors and thus {\it create} Majorana fermions. 
Indeed, we find that, for long disordered wires, the total area of the topological
phase is conserved. Strikingly, if the scattering is regular e.g. due to a superlattice, the area of
the topological phase can be made to increase beyond the clean
value, raising the possibility to further engineer topological order.

This article is organized as follows: we develop a theory
capable of studying topological phase transitions in the presence of
individual (possibly random) potential
configurations, rather than calculating average quantities.
First, we focus on the almost depleted wire and recover in the weak-disorder limit
the earlier results of Refs.~\cite{Motrunich01,Piet11}, namely that disorder
is {\it always} detrimental to the topological order for p-wave superconductors.
%Next, we focus on the almost depleted wire and recover the earlier results of
%Refs.~\cite{Motrunich01,Piet11} for average phase
%boundary for the effective p-wave model. In particular, we recover the result that the disorder is
%{\it always} detrimental to the topological order for p-wave
%superconductors.
We then show how, for individual disorder configurations, one
can relate the phase diagram to an experimentally accessible quantity:
the normal state conductance. This result allows us to solve {\it inter alia} the
Gaussian disordered p-wave problem exactly for all values of the
disorder strength (Fig.~\ref{fig_pwave}).
Finally, we focus on the experimentally relevant case of a
semiconductor nanowire with 
s-wave superconductivity.  We find that, unlike its 
p-wave counterpart, topological s-wave superconductivity is resistant to
disorder 
%and the conditions (i) and (ii) are in
%fact not essential for the survival of Majorana fermions 
(Fig.~\ref{FIG:swave}). 

\section{Spinless p-wave superconducting wire} 

We start with the spinless p-wave Hamiltonian, as the calculation is easier to follow
and illustrates the essential concepts. 
% Moreover, in the large
% $B$ limit, the s-wave Hamiltonian reduces to the p-wave
% case, allowing for further checks on its validity.
We note that the disordered p-wave model was solved at half-filling
as well as for specific position-dependent potentials~\cite{Lang12,DeGottardi12}. 
Here, we present a general solution.

The Bogoliubov-de~Gennes (BdG) Hamiltonian for a spinless $p$-wave
superconductor in one dimension is:
\begin{equation}
H= h(p,x)\tau_z + \vdelta\,p \, \tau_x,
\end{equation}
where $h(p,x)=p^2/2m+V(x)-\mu$ is the (spinless) 
single-particle Hamiltonian, $p$ is the momentum operator,
$m$ the electron mass, $V(x)$ the single-particle potential,
$\mu$ the chemical potential, and $\vdelta\,p$ the p-wave pairing operator.
Here and below $\tau_i$
($i=x,y,z$) denote the Pauli matrices in the electron-hole space.
In order to make use of the chiral symmetry of the Hamiltonian,
we apply a unitary transformation with $U = \exp(i\tau_x \pi/4)$, 
that casts the Hamiltonian into an off-diagonal form.
We note that similar argumentation was used to study zero modes in 
d-wave superconductors~\cite{inanc99}.
%The main use of this form is that 
% It is now easy to see that 
The energy $E=0$ Majorana fermion solutions
are then either of the form $\chi_+=\bigl({\varphi_+\atop{0}}\bigr)$ or of the form
$\chi_-=\bigl({0\atop{\varphi_-}}\bigr)$, with $\varphi_\pm={\rm e}^{\pm k_\vdelta x}\psi$,
% $\left(h(p,x) \pm  i u p \right)\varphi_{\pm}=0$.
% The $p$-linear term can be removed by the transformation
% $\varphi_\pm={\rm e}^{\pm k_\vdelta x}\psi$, 
where $k_\vdelta=m\vdelta/\hbar$ and
%We then find that 
$\psi$ locally satisfies the normal state equation
% \begin{equation}
% \left(-\frac{\hbar^2}{2m}\partial_x^2+V(x)-\mu+\frac{\hbar^2k_\vdelta^2}{2m}\right)\psi=0.
% \label{EQ:Eff_SE}
% \end{equation}
% \color{red}
% $\partial_x^2 \psi =
% 2m\hbar^{-2}\big(V(x)-\bar{\mu}\big)\psi,$
% \color{black}
$h(p,x)\psi=-(\hbar^2 k_\vdelta^2/2m)\psi$.
%We identify this equation as the normal state equation with an effective chemical potential
%where $\bar{\mu} = \mu - \hbar^2 k_\vdelta^2/2m$. 
However, 
%note that
it is $\varphi_\pm$
that needs to be normalized, rather than $\psi$ itself.
Hence a diverging solution $\psi$
as $x\rightarrow \pm \infty$
%leads to normalizable $\chi_\pm$,
is permissible if the divergence is not faster than ${\rm e}^{\pm k_\vdelta x}$.

We now construct the Majorana fermion state. For the sake of
concreteness, we consider an interface between an half infinite ($x>0$)
wire, with the vacuum $x<0$ (a normal insulator)
implemented via
the boundary condition (BC) $\chi(0)=0$. We note that, it is easy to generalize to BCs of the form
$a \chi(x_0) +b \frac{d\chi}{dx}\vert_{x_0}=0$. We also require $\chi \rightarrow 0$ 
sufficiently fast as $x\rightarrow \infty$ to ensure normalizability. Then, choosing
$\psi = g(0) f(x) - f(0) g(x)$, with $f$ and $g$ the local solutions
of the normal state equation, ensures that
$\chi$ fulfills the BC at $x=0$. 
We focus on solutions
that
behave as 
$\psi \sim e^{\Lambda x}a(x)$ for large $x$, with $a(x)$ a
nondivergent function and
$\Lambda(\bar{\mu})$ a real function of $\bar{\mu} = \mu - \hbar^2 k_\vdelta^2/2m$. 
For solutions
that diverge or 
%go to zero 
decay
faster (slower) than $e^{\Lambda x}$ we set $\Lambda={\rm sgn}(\Lambda) \times\infty$ ($\Lambda=0$).
We identify three cases
(i) $\Lambda<-k_\vdelta$, (ii) $|\Lambda|<k_\vdelta$, and (iii)~$k_\vdelta<\Lambda$.
For case (i)~$\psi$ is a bound normal state solution that fulfills both BCs and there are two zero modes $\chi_+$ and
$\chi_-$. Under a small perturbation, $\psi$ no longer satisfies the BCs, and hence the two
solutions $\chi_\pm$ will shift away from $E=0$, 
i.e.~they are not topologically protected.
This corresponds to an accidental level crossing at $E=0$~\cite{footnote_accidental}.
In case (ii)~there is only one state, $\chi_-$,
%which is 
the topologically protected Majorana state,
and in case (iii)~there are no
 zero modes and thus no 
Majorana state. 
We thus obtain a formula for the topological charge:
\begin{equation}
{\cal Q}={\rm sgn}\left(\left|\Lambda(\mu-\hbar^2 k_\vdelta^2/2m)\right|/k_\vdelta-1\right),
\label{EQ:TOP_COND}
\end{equation}
where ${\cal Q}=-1$ means the wire is topological.
This is the central result for the $p-$wave part of our work.

The topological robustness of the zero energy solutions is due to the
fact that only the asymptotic limit of the solution $\psi$ of the 
effective Schr\"odinger equation matters
for its existence. Any local perturbation
(unless infinite) cannot change this asymptotic limit.

%In order to see the topological robustness
%of the zero energy solutions,  first note that it is only the
%asymptotic limit of the solutions
%$\psi$ of the effective Schr\"odinger equation that matters for the
%existence of the solutions. Next, notice that local perturbations of the potential (unless infinite)
%cannot change the asymptotic behavior of the solutions regardless of
%their size and shape. Thus if there is a zero mode of the BdG
%hamiltonian for some potential profile (i.e.it is in the
%topological state) so will any other Hamiltonian that differs from the
%former by a local perturbation, demonstrating topological invariance.

\begin{figure}
\includegraphics[width=\linewidth]{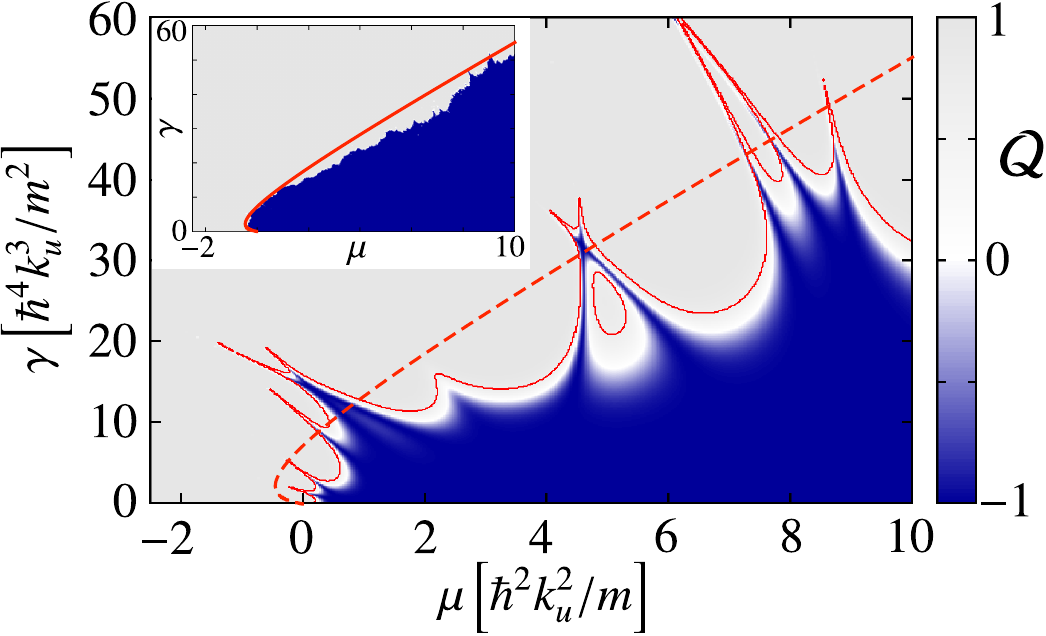}
\caption{Topological charge 
%$\mathcal{Q}=\det(r)$ 
of a disordered p-wave wire as a function
of 
%chemical potential 
$\mu$ and disorder strength $\gamma$, for a single disorder configuration
in a short wire ($L=100a$, with $a$ the lattice constant).
The inset shows a single disorder configuration in a long wire ($L=10000a$).
The red solid line is the phase boundary computed from
Eqs.~\eqref{EQ:TOP_COND} and the normal state conductance $G$, the red
dashed lines are
from Eqs.~\eqref{EQ:TOP_COND} and \eqref{EQ:LYEXP}. The numerical
calculation was done in a TB model with $k_u=10a^{-1}$ and a chemical potential in the leads
$\mu_\text{lead}=0.5 \hbar^2/2ma^2$. }\label{fig_pwave}
\end{figure}

For a disordered (normal-state) wire, $\Lambda$ is called the Lyapunov
exponent and can be estimated from the conductance as:
$\Lambda=-\log (G/G_0)/2L,$
where $L$ is the wire length and $G_0$ the conductance quantum~\cite{RMT_Carlo}. Hence,
for fixed $u$, Eq.~\eqref{EQ:TOP_COND} 
%allows one to 
determines 
the topological charge of a p-wave quantum wire from its normal state
conductance alone. In short wires, $\Lambda$ fluctuates strongly
as the chemical potential varies, 
%and as a consequence there are 
leading to
multiple changes of the
topological phase. This is shown on the example of a single disorder realization in a short wire in
Fig.~\ref{fig_pwave}, where we computed the topological charge within a tight-binding (TB) model from $Q=\det(r)$ where
$r$ is the reflection matrix~\cite{Akhmerov11}. The numerical computation was performed using the Kwant code~\cite{Kwant}.
The topological phase boundary computed from Eq.~\eqref{EQ:TOP_COND} and the
numerically computed normal state conductance agrees very well with the $\det(r)$-criterion; small
deviations of the exact position of the phase boundary are due to finite size effects.

For longer wires the Lyapunov exponent is a self averaging quantity,
i.e.~$\Lambda(L)\rightarrow \bar{\Lambda}$, as
$L\rightarrow\infty$, where
$\bar{\Lambda}$ is the average Lyapunov exponent~\cite{RMT_Carlo}. For
a wire with Gaussian disorder $\langle V(x)V(y)\rangle=\gamma \delta(x-y)$ at energy $\epsilon$, it can be
obtained in closed form~\cite{Halperin65, ItzDrou}:
\begin{subequations}
\begin{eqnarray}
\!\!\!\!\bar{\Lambda}(\epsilon)\!\!&=&\!\!
\frac {m^{1/2}}{\hbar\lambda}
F\big(\lambda^2 \epsilon\big),\quad~~
\lambda=\left(\frac{\hbar}{\gamma m^{1/2}}\right)^{1/3},
\\
\!\!\!\!F(x)\!\!&=&\!\!-\frac{1}{2}\frac{d \ln \Big({\rm Ai}\big(\!-\!2^{1/3}x\big)^2+{\rm Bi}\big(\!-\!2^{1/3}x\big)^2\Big)}{dx}.
\end{eqnarray}%
\label{EQ:LYEXP}
\end{subequations}%
Then the topological transition condition Eq.~(\ref{EQ:TOP_COND}) becomes
$\hbar|\bar{\Lambda}(\mu-m\vdelta^2/2)|=m \vdelta$, valid
for the entire range of $\mu$, $\vdelta$, $\gamma$ and shown as a red
dashed line in
Fig.~\ref{fig_pwave} and its inset. The inset also shows numerics for a {\it single} disorder
configuration for a long wire, demonstrating that due to the self-averaging long wires
have a well-defined universal topological phase (similar numerics, as well as an argument for
weak disorder
was presented in~\cite{Pientka12}). At high energies, we have
the golden rule result $\Lambda\sim1/4\ell_\text{tr}$, where $\ell_\text{tr}=\hbar^2 (\mu-m\vdelta^2/2)/\gamma m$ is
the transport mean free path,
and find 
a topological transition 
at $k_\vdelta \ell_\text{tr}=1/4$, in agreement with Ref.~\cite{Piet11, footnotefactor2}.

From Eq.~\eqref{EQ:TOP_COND} it can be also concluded that for $\bar{\mu}>0$ any scattering is
detrimental to the topological phase: Then $\Lambda=0$ in the clean
system 
% (the normal state solutions are extended),
and any scattering leads to $\Lambda\geq 0$. For $\bar{\mu}<0$,
potential fluctuations generate islands of topological
regions which may hybridize to 
induce a topological state as seen in the inset of Fig.~\ref{fig_pwave}. However, this is a relatively small effect.
We shall see below this picture is drastically different for the experimentally relevant proximity nanowire systems.

\section{Rashba wire in proximity to an s-wave superconductor} 

We now focus on the experimentally more relevant system: a nanowire
with Rashba spin-orbit coupling (SOC) in proximity
to an s-wave superconductor. The BdG Hamiltonian is then given as~\cite{Lutchyn10, Oreg10}:
\begin{equation}\label{EQ:swave_ham}
H=h(p,x)\tau_z+\alpha p \sigma_y \tau_z + B \sigma_x + \Delta \tau_x,
\end{equation}
where $h(p,x)=p^2/2m + V(x)-\mu$ is the (spinless) single-particle Hamiltonian,
$\alpha$ the SOC strength,
$B$ the Zeeman splitting and $\Delta$ the induced s-wave order parameter. $\sigma_i$ ($i=x,y,z$) are the Pauli
matrices in spin space. The topological state appears for $B^2 > \Delta^2 +\mu^2$.
In this single orbital mode limit, the system is in class BDI, which is
distinguished from class D by the presence of the chiral symmetry.
This allows to bring the Hamiltonian into off-diagonal form~\cite{Tewari12}, and
a solution can be found in a similar spirit to the p-wave case
considered above (details of the calculation can be found in the Appendix). In particular, the zero-energy Majorana states are of again of the form
$\chi_+=\bigl({\phi_+\atop{0}}\bigr)$ or
$\chi_-=\bigl({0\atop{\phi_-}}\bigr)$, but it the present case $\phi_\pm$ is a spinor
satisfying a $2\times 2$ nonhermitian eigenvalue problem:
\begin{equation}\label{EQ:swave_nonhermitian}
(h(p,x)\sigma_z \pm B \pm \Delta\sigma_x -i\alpha p \sigma_x)\phi_{\pm}=0
\end{equation}
Zero-energy solutions of this equations can be found in closed form only for 
small $\alpha$, but larger values of SOC
do not change the qualitative picture, but rather renormalize the
topological-normal phase boundaries. To order $\alpha^2$ the solution reads
\begin{eqnarray}
&&\phi_\pm=\xi_\pm(\epsilon) {\rm e}^{\pm \kappa x} \big(A f(x;\epsilon) + B g(x;\epsilon) \big) \nonumber \\
&& \quad\quad+ \xi_\pm(-\epsilon) {\rm e}^{\mp \kappa x} \big( C
f(x;-\epsilon) + D g(x;-\epsilon) \big),
\label{EQ:MajoranaWF}
\end{eqnarray}
where $\epsilon=\sqrt{B^2-\Delta^2}$, $\kappa=m\alpha \Delta/\hbar\epsilon$, and
$\xi_+(\epsilon)$ is the eigenvectors of the $2\times 2$ matrix
$\epsilon \sigma_z + \Delta \sigma_x$ with positive eigenvalue. $f(x; \epsilon)$
and $g(x;\epsilon)$ are, as 
%in the p-wave case, 
above, the two linearly independent
solutions of $h(p,x) \psi = \epsilon \psi$, with $f$ decaying and $g$ increasing.
%Then, $\phi_\pm$ is a valid zero-energy solution (and thus a Majorana fermion) if it is normalizable and satisfies the BCs.
Then, $\phi_\pm$ is a zero-energy Majorana state if it is normalizable and satisfies the BCs.

We assume again without loss of generality that the system is in a
normal insulator state for $x<0$ and the BC
$\phi(0)=0$.
We identify three cases:
(i)~If $B>\Delta$, and $|\Lambda(\mu\pm\epsilon)|<|\kappa|$ or $|\Lambda(\mu\pm\epsilon)|>|\kappa|$, there are
two decaying and two diverging solutions and the BC
at $x=0$ can only be satisfied accidentally, namely if $f(0,\pm\epsilon)=0$. Then there is
also a second solution in the other sector, and the zero-energy states are not protected. The system is
thus in the trivial state with the possibility of accidental zero modes.
(ii)~If $B<\Delta$, then both  $\kappa $ and $\epsilon$ are imaginary, hence there are always
two decaying and two diverging solutions. However, there are no accidental zero modes with
$f(\pm\epsilon)$ already fulfilling the BC because this would mean $f$ is an eigenfunction
of (Hermitian) $h$ with an imaginary eigenvalue.
(iii)~If $B>\Delta$ and
$\vert\Lambda(\mu\pm\epsilon)\vert<\vert\kappa\vert<\vert\Lambda(\mu\mp\epsilon)\vert$, there are one diverging and three decaying
solutions in one sector and one decaying and three diverging solutions
in the other sector. Then the BC at $x=0$ can be generally
satisfied in the sector that has three decaying solutions and there is
a Majorana state. As before, the solution is
robust, because local perturbations do not change the asymptotic behavior of
$f$ and $g$.
In summary we have:
\begin{equation}
{\cal Q}= {\rm sgn}\left(
\frac{|\Lambda(\mu+\epsilon)|}{m\alpha\Delta/\hbar\epsilon}-1
\right)\,
{\rm sgn}\left(
\frac{|\Lambda(\mu-\epsilon)| }{m\alpha\Delta/\hbar\epsilon} -1
\right).
\label{EQ:TOP_CHARGE}
\end{equation}
This is our central formula for the s-wave case. The first term in
Eq.~\eqref{EQ:TOP_CHARGE} reduces to Eq.~\eqref{EQ:TOP_COND} in the
large $B$-limit (i.e.~only the ``spin-down'' band is contributing), recovering the p-wave result, while
the second term is due to the presence of the
``spin-up'' band and introduces new physics.
 In summary, a transport gap in one of the ``spin-bands''
induces topology in the other ``spin-band'', in contrast to the clean case
where one spin-band is removed by a spectral (Zeeman) gap.

\begin{figure}
\includegraphics[width=\columnwidth]{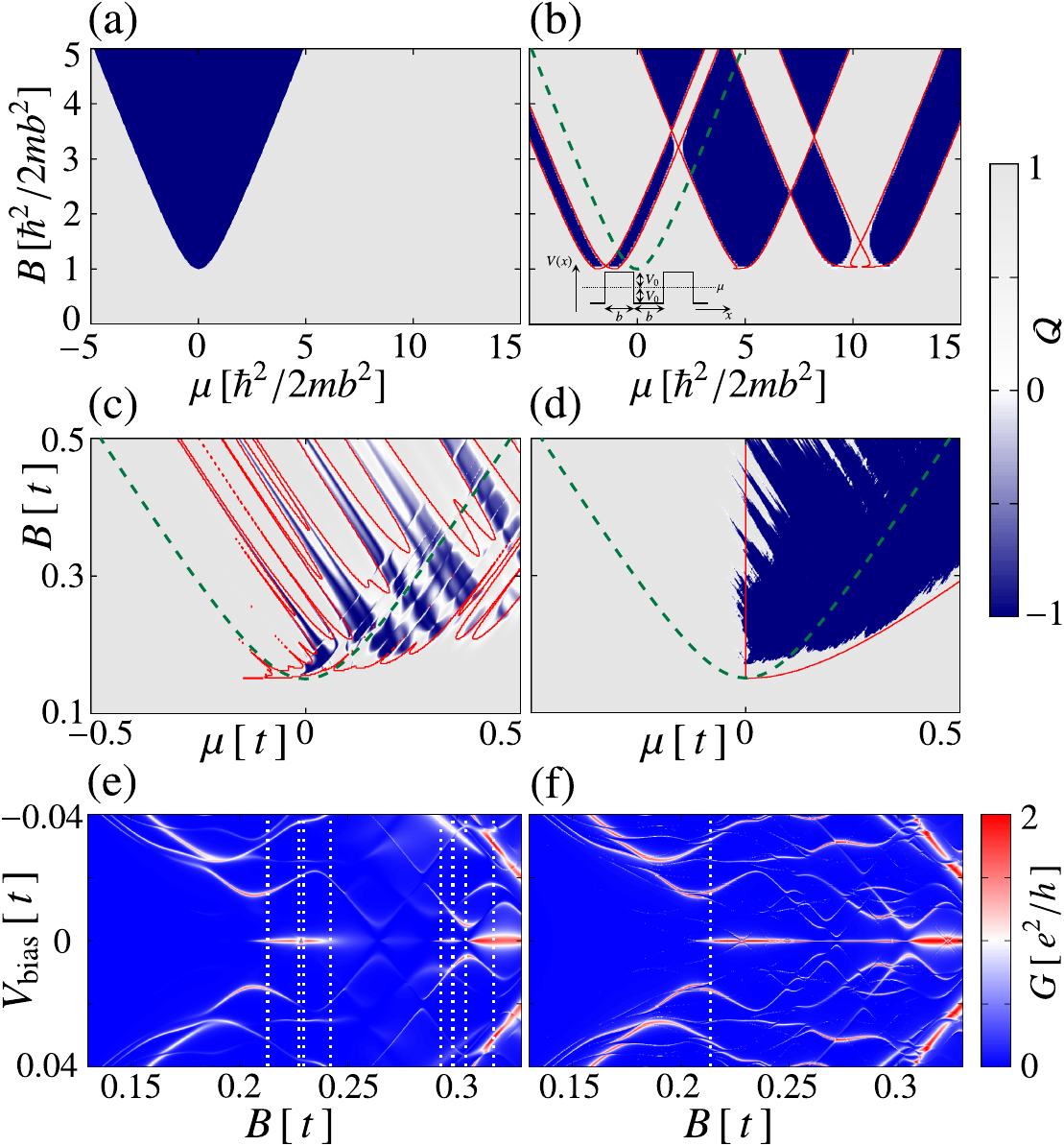}
\caption{Topological charge 
%$\mathcal{Q}=\det(r)$ 
as a function of 
chemical potential
$\mu$ and Zeeman splitting $B$ for a (a) clean system, (b) a superlattice, and (c, d) disorder.
Red lines in (b-d) are phase boundaries calculated from Eq.~\eqref{EQ:TOP_CHARGE}, green dashed lines
show the clean phase boundary for comparison.
(b) The superlattice (see inset) parameters were $d=3b$, $V_0=8\hbar^2/2mb^2$,
$\Delta=\hbar^2/2mb^2$, and $k_\text{so}=0.05 b^{-1}$ with $k_\text{so}=m\alpha/\hbar$, and
the numerical calculation was done using a transfer matrix method in Mathematica.
The numerical calculations in (c-f) were done within a TB model:
(c) shows 
%the topological charge 
$\mathcal{Q}$
for a single disorder realisation in a short ($L=100a$ with $a$ the lattice constant)
and (d) in a long ($L=4000a$) wire, (e) and (f) the respective tunnel conductances for a fixed $\mu=0.3 t$,
with $t=\hbar^2/2ma^2$. White dashed lines in (e, f) indicate the
boundaries of the topological phase in (c, d). The remaining TB parameters were $k_\text{so}=0.05 a^{-1}$,
$\Delta=0.15 t$, $\gamma=0.06 t^2$ , and the chemical potential in the leads
$\mu_\text{leads}=0.5 t$. For the tunneling conductance in (e, f) a barrier of height
$1.5 t$ was added on one lattice site next to one end of the wire.
\label{FIG:swave}
}
\end{figure}

We now apply our formula Eq.~(\ref{EQ:TOP_CHARGE}) to the case of
regular scattering (i.e.~from a superlattice). 
%In the clean case, 
For a clean wire,
the required odd number of channels for the topological state is
only achieved if the chemical potential is within the Zeeman gap~\cite{Lutchyn10, Oreg10}.
% and hence close to the band bottom 
The perfect backscattering from a
superlattice (or equivalently, minigaps) allow this for a larger range
of $\mu$.
%chemical potentials. 
% 
Strikingly, even a superlattice formed from topologically trivial
pieces can be topological. In summary, regular scattering can induce topological
order out of the Zeeman gap,
%away from the band bottom, 
enlarging the topological phase area beyond its clean wire value,
as shown in Fig.~\ref{FIG:swave}(a,b).

%-------------
%-------------

In the experimentally relevant case of irregular scattering, we use the average Lyapunov exponent given by Eq.~(\ref{EQ:LYEXP}) to
determine the (not-averaged) phase boundary of a long quantum wire from Eq.~\eqref{EQ:TOP_CHARGE}.
Noting that $\bar{\Lambda}$ is a monotonous function of energy, we get:
\begin{equation}
\mu_\pm=F^{-1}(m^{1/2}\lambda\alpha\Delta/\sqrt{B^2-\Delta^2})/\lambda^2\pm\sqrt{B^2-\Delta^2}
\label{EQ:long_PB}
\end{equation}
In the clean limit, $\lambda\rightarrow\infty$, we recover the
ballistic result: $\mu_\pm=\pm\sqrt{B^2-\Delta^2}$.
In contrast to the common wisdom based on the effective p-wave model, we find that the topological region is not
destroyed by disorder but merely shifted to higher chemical potentials. In fact the chemical potential (or gate)
range where the wire is topological,
$\mu_+-\mu_-=2\sqrt{B^2-\Delta^2}$, is {\it independent} of the disorder strength. Thus the total area of the topological
region in the $(B,\mu)$ plane is conserved.  We stress that this result is valid to all orders in disorder strength.

This picture is confirmed numerically in Fig.~\ref{FIG:swave}(d), where we compare our theoretical prediction
Eq.~(\ref{EQ:long_PB}) with our numerical results for a long,
disordered nanowire. We observe that the disorder creates
a well-defined topological region for a parameter range where the
clean wire is trivial. 
In a short wire, the topological
phase, plotted in Fig.~\ref{FIG:swave}(c), is more fragmented due to the fluctuations in the normal state conductance in agreement
with Eq.~\eqref{EQ:TOP_CHARGE}.
Nevertheless, a clear Majorana ZBP appears in the
tunneling conductance for both wires,
as shown in Fig.~\ref{FIG:swave}(e,f).
Note that
the clean wire would have been in the trivial phase 
% without Majorana fermions 
% without disorder 
for the range of
parameters shown in Fig.~\ref{FIG:swave}(e,f). 

\section{discussion} 

Recently, it was argued that ZBPs in nanowires
may appear even without Majorana fermions~\cite{Liu12, Bagrets12, Pikulin12, Rainis13}. Here we caution against this interpretation.
A ZBP out of the clean topological phase boundary may well be a Majorana fermion within the dirty
topological phase boundary, especially if $B>\Delta$ and the ZBP
remains for a range of magnetic field~\cite{B-dep}.
In fact, we note that the nanowires in Ref.~\cite{mourik12} have lengths of the order of several $\ell$ in their normal state, 
and hence we expect the process of disorder-induced-topology discussed
here to play a role. 
The lowering of the threshold magnetic field for Majorana fermions with disorder reduces the necessity to fine-tune the chemical
potential. Moreover, the requirement of quasiballistic wires is also relaxed, 
possibly explaining why Majorana fermions were routinely
observed on several samples. The 
experiments of Ref.~\cite{mourik12} are in the limit of short wires where the Majorana ZBP in a disordered nanowire 
vanishes and reappears repeatedly due to the fragmentation of the
topological phase (see Fig.~\ref{FIG:swave}(e) and the Appendix). 
Such multiple disappearances and reappearances of the ZBP with increasing magnetic field
have been observed experimentally (see Supporting Online Material of \cite{mourik12}), supporting the picture presented in this work.
This reentrant ZBP is due to a repeated change from topological to trivial phase and vice versa,
in contrast to the Majorana oscillations discussed in \cite{Rainis13, DasSarma12} where the
wire is always topological.

In conclusion, we studied the effects of scattering from a potential
in one-dimensional topological superconductors. We obtained analytical
formulas for the phase boundaries in the case of regular and irregular
scattering, valid to all orders in the potential strength and 
applicable also to single potential configurations. Our main result is
that disorder does {\it not} always destroy topological order, contrary
to expectations from p-wave models:
for proximity-coupled nanowires the phase merely shifts to larger chemical
potential, conserving the total area. With a periodic potential
modulation the phase area can further be increased.

%In conclusion, we studied the effects of single-particle scattering on the topological properties of a quantum wire in contact with an
%s-wave superconductor. 
%We obtained analytical formulas for the phase boundaries valid for regular and irregular scattering. In the case
%of irregular scattering, our formulas apply to all orders in the disorder strength as well as to individual members of the ensemble.
%Our
%main result is that disorder does {\it not} always destroy topological order, contrary to the general expectation based on effective models,
%rather the topological phase is merely shifted to higher chemical potentials, while the total phase area is conserved.
%Moreover, one can further increase the topological phase area with a periodic modulation of the gate potential.

We acknowledge discussions with C.W.J. Beenakker.
This work was supported by funds of the Erdal \.{I}n\"{o}n\"{u}
chair, TUBITAK under grant No. 110T841, TUBA-GEBIP,
and an ERC Advanced Investigator grant.

\appendix

\section{Results for parameters as in the Delft experiment}

\begin{figure}
\includegraphics[width=\columnwidth]{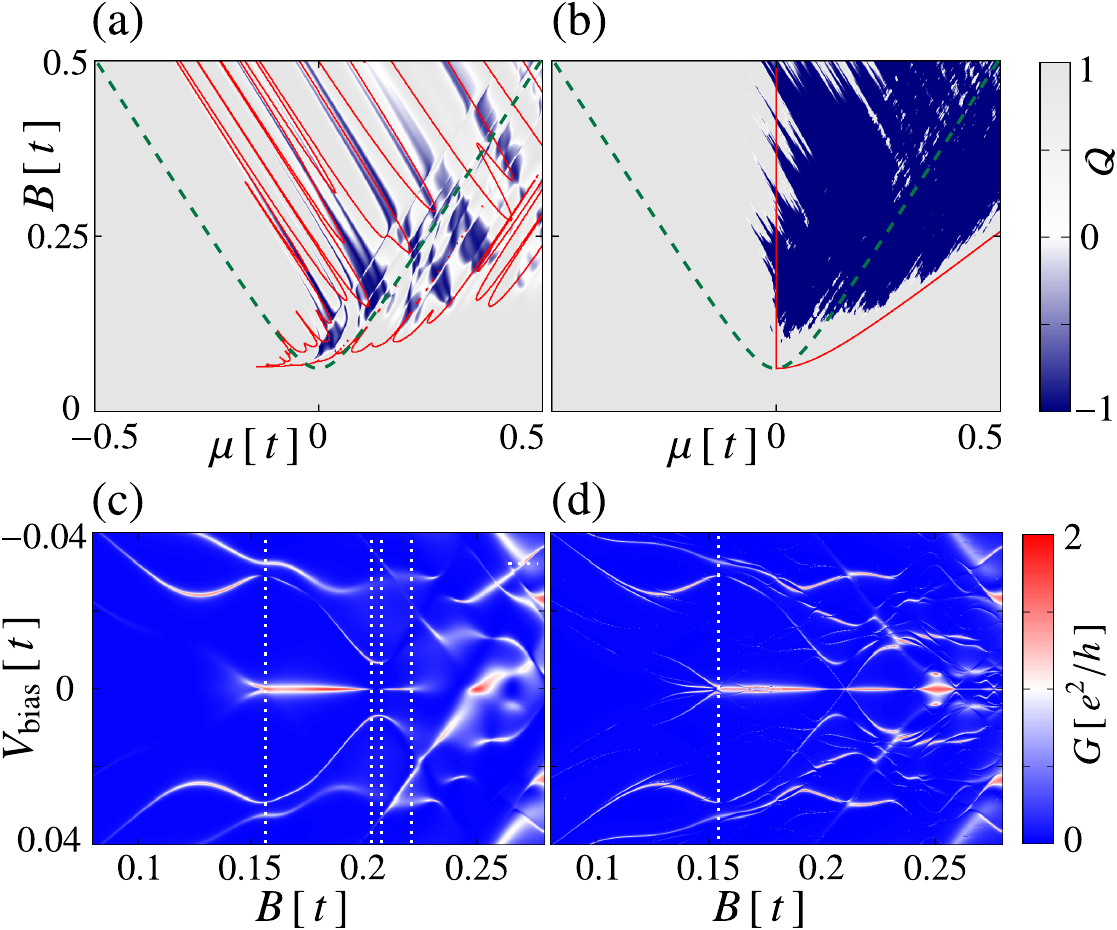}
\caption{
(a, b) Topological charge $\mathcal{Q}=\det(r)$ as a function of chemical potential
$\mu$ and Zeeman splitting $B$ for a (a) short ($L=100a$) and a (b) long ($L=4000a$)
disordered nanowire. Red lines are phase boundaries calculated from Eq.~(8) in the main text, green dashed lines
show the clean phase boundary for comparison.(c, d) The corresponding tunnel conductance 
for fixed $\mu=t$. Parameters in the tight-binding model were chosen to fit the Delft experiment
\cite{mourik12}: Choosing the lattice spacing $a=25$nm,
we obtain a hopping constant $t=\hbar^2/2 m a^2=4.064$meV for
the effective mass $m=0.015 m_\text{e}$ of InSb. The short wire with $L=2.5\mu$m then corresponds
to the experimental situation. The other parameters used in the 
simulation were $l_\text{so}=1/k_\text{so}=10 a=250$nm,
$\Delta=0.0615 t=250\mu$eV, $\gamma=0.0492 t^2$, and the chemical potential in the leads
$\mu_\text{leads}=0.5 t$. For the tunneling conductance in (c, d) a barrier of height
$1.5 t$ was added on one lattice site next to one end of the wire.}
\label{fig:delft_exp}
\end{figure}

Fig.~\ref{fig:delft_exp} shows the results of a numerical simulations for parameters
applicable to the Delft experiment \cite{mourik12}. The experiment
is in the regime of intermediate spin-orbit coupling strength. As a 
consequence, there is some deviation between the analytical solution
obtained in the weak spin-orbit limit and the numerical results. Still,
all of the characteristic features discussed in the main text are present:
The creation of topological phases outside the clean phase boundaries,
the lowering of the threshold magnetic field $B$ for entering the
topological phase with disorder, the conservation
of the area of the topological phase in a long wire, and the repeated appearance and
disappearance of the Majorana peak in the short wire limit (which is the
experimentally relevant situation).

\section{Details of the calculation of Eq.~(7)}

The chiral symmetry of the Hamiltonian (5) in the main text implies that there is an
operator that anti-commutes with the Hamiltonian: 
$\sigma_y \tau_y$. In the basis that diagonalizes
this operator with degenerate blocks off-diagonalizes the Hamiltonian.
In particular, we find that $U=(1+i\sigma_x)(1+i\tau_x)\big((1+\sigma_z)+(1-\sigma_z)\tau_x\big)/4$
transforms the Hamiltonian to:
\begin{equation}
H=h(p,x)\sigma_z\tau_y-\alpha p \tau_y + B \sigma_x \tau_x + \Delta \tau_x.
\end{equation}
Then the zero energy Majorana states are of either of the form
$\chi_+=\bigl({\phi_+\atop{0}}\bigr)$ or
$\chi_-=\bigl({0\atop{\phi_-}}\bigr)$, where $\phi_\pm$ satisfy
a $2\times 2$ nonhermitian eigenvalue problem with eigenvalue zero:
\begin{equation}
(\mp i h(p,x)\sigma_z \pm i\alpha p + B \sigma_x + \Delta)\phi_{\pm}=0.
\end{equation}
After performing a rotation in
$\sigma$ space around the $x$-axis that transforms
$\sigma_z\rightarrow \sigma_y$ and premultipling with $\pm \sigma_x$ 
we obtain Eq.~(6) of the main text:
\begin{equation}
\tilde{H} \phi_\pm=(h(p,x)\sigma_z \mp B \mp \Delta\sigma_x -i\alpha p \sigma_x)\phi_{\pm}=0
\end{equation}

We now construct the zero energy solution for small $\alpha$. 
First, we perform an imaginary gauge transformation: $\phi\rightarrow
e^{-\kappa_\alpha x} \phi$, where $\kappa_\alpha$ is an order $\alpha$ parameter
that is yet to be determined. Then we have $p\rightarrow
p+i\hbar\kappa_\alpha$. Next, we collect terms of order $\alpha$ and treat
them as perturbations. We then have $\tilde{H}=H_0 + H_1$ with
\begin{subequations}
\begin{eqnarray}
&&\!\!\!\!\!\! H_0= h(p,x)\sigma_z \mp B \mp \Delta\sigma_x \\
&&\!\!\!\!\!\! H_1= -i\alpha p \sigma_x +i\frac{\hbar\kappa p}{m} \sigma_z +
\hbar\kappa \alpha \sigma_x-\frac{\hbar^2\kappa^2}{2m} \sigma_z\,.
\end{eqnarray}
\end{subequations}
The last two terms can be absorbed into $H_0$ by redefining $\mu$ and $\Delta$
and will be neglected in the following.

Zero-energy solutions of $H_0$ are of the form $\xi_\pm(\epsilon) \psi(x;\epsilon)$ where
$h(p,x) \psi(x;\epsilon) = \epsilon \psi(x;\epsilon)$, $\xi_\pm(\epsilon)$ are
the eigenvectors  of the $2\times 2$ matrix $\epsilon \sigma_z \mp \Delta \sigma_x$
with eigenvalue $\pm \sqrt{\epsilon^2 + \Delta^2}$, and $\epsilon=\sqrt{B^2-\Delta^2}$.
$\psi(x; \epsilon)$ can be again written as a linear combination of two independent
solutions $A f(x; \epsilon) + B g(x; \epsilon)$ where we choose $f$ to be decaying
and $g$ increasing.

We now choose $\kappa_\alpha=\mp m \alpha \Delta/\hbar \epsilon$
such that $H_1$ anticommutes with $\epsilon \sigma_z \mp \Delta \sigma_x$.
Then, $H_1$ is off-diagonal in the basis of $\xi_\pm(x; \epsilon)$ and thus
the contribution of $H_1$ vanishes to first order in perturbation theory. Hence,
\begin{eqnarray}
&&\phi_\pm=\xi_\pm(\epsilon) {\rm e}^{\pm \kappa x} \big(A f(x;\epsilon) + B g(x;\epsilon) \big) \nonumber \\
&& \quad\quad+ \xi_\pm(-\epsilon) {\rm e}^{\mp \kappa x} \big( C
f(x;-\epsilon) + D g(x;-\epsilon) \big),
\end{eqnarray}
is a zero-energy solution up to order $\alpha^2$ with $\kappa = m \alpha \Delta/\hbar \epsilon$.


\begin{thebibliography}{9}
\bibitem{footnote_accidental} Thus an unprotected
pair of zero-energy modes is always accompanied by a
state at the Fermi energy for the normal system,
without superconductivity.
\bibitem{footnotefactor2}
In \cite{Piet11}, the criterion is $k_\vdelta \ell=1/2$, with
$\ell = 2 \ell_\text{tr}$.
\end{thebibliography}
\end{document}